\documentclass[a4paper,12pt]{article}
\usepackage{jheppub}
\usepackage{amsmath,amssymb,bm,mathrsfs}
\usepackage{mathtools,mathptmx,array,slashed}
\usepackage{graphicx,graphics,subfigure}
\usepackage{ctable,tabularx,multirow}
\usepackage[colorlinks=true,linkcolor=blue,citecolor=blue,urlcolor=blue]{hyperref}

\def\be{\begin{equation}}
\def\ee{\end{equation}}
\def\bea{\begin{eqnarray}}
\def\eea{\end{eqnarray}}
\def\ba{\begin{aligned}}
\def\ea{\end{aligned}}
\def\nn{\nonumber}
\def\p{\partial}

\title{Classifying topology of consistent thermodynamics of the four-dimensional
neutral Lorentzian NUT-charged spacetimes}

\author{Di Wu}

\affiliation{School of Physics and Astronomy, China West Normal University,
Nanchong, Sichuan 637002, People's Republic of China}

\emailAdd{wdcwnu@163.com}

\abstract{In this paper, via employing the uniformly modified form of the generalized
off-shell Helmholtz free energy, we investigate the topological numbers for the
four-dimensional neutral Lorentzian Taub-NUT, Taub-NUT-AdS and Kerr-NUT spacetimes,
and find that these solutions can also be classified into one of three types of those
well-known black hole solutions, which implies that these spacetimes should be viewed
as generic black holes from the viewpoint of the thermodynamic topological approach.}


\arxivnumber{2302.01100v4}

\begin{document}

\maketitle

\flushbottom

\section{Introduction}

In the great family of four-dimensional exact solutions in General Relativity, the Lorentzian
NUT-charged spacetimes belong to a very important class of asymptotically locally flat solutions
to the Einstein field equations \cite{AM53-472,JMP4-915}. Since its birth, the prevailing view
has commonly excluded the Lorenzian NUT-charged spacetimes from the big family of black holes,
due to the Misner's identification of the NUT charge parameter as unphysical \cite{JMP4-924}.
However, in recent years, there has been a surge in the studies that exploit the Lorentzian
NUT-charged spacetimes as black holes to explore their physical properties, including thermodynamics
\cite{PRD100-064055,CQG36-194001,JHEP0719119,PLB798-134972,JHEP0520084,PRD100-104016,PLB832-137264,
PLB802-135270,IJMPD31-2250021,JHEP0821152,PRD101-124011,PRD105-124034,PRD100-101501,PRD105-124013,
2209.01757,2210.17504,PRD103-024052,JHEP0321039,PRD106-024022,EPJP130-124,JHEP1022044,JHEP1022174},
geodesic motion \cite{EPJC80-1000}, Kerr/CFT correspondence \cite{EPJP134-580,PLB820-136568},
black hole shadow \cite{PLB816-136213}, weak cosmic censorship conjecture \cite{PRD101-064048,
PRD103-106011}, holographic complexity \cite{PRD100-066007}, and so on \cite{EPJC79-466,
SCPMA64-260411,EPJC79-161,PRD102-124058,PRD103-126012,EPJC81-621,EPJC81-838}. Thus, a question
arises naturally as to whether the Lorentzian NUT-charged spacetimes are generic black holes.

On the other hand, the Lorentzian NUT-charged spacetimes have many peculiar properties that are
mainly due to the presence of the wire/line singularities at the polar axes $\theta = 0$ and
$\theta = \pi$), which are often called as the Misner strings. For instance: (I) The NUT charge
has many different meanings and interpretations \cite{GRG5-603,JMP22-2612,NPB289-735,NPB311-739,
RMP70-427}, and there are many different explanations of the physical origin of the NUT-charged
spacetimes \cite{PCPS66-145,PCPS70-89}. Up to date, there is no unified explanation for these
facts; (II) The Taub-NUT de Sitter spacetimes not only provide counterexamples to the maximal
mass conjecture but also violate the entropic $N$-bound \cite{PRL91-061301,NPB674-329,
IJMPA19-3987}; (III) The thermodynamic volume of the Euclidean Taub-NUT-AdS spacetime
can be negative \cite{CQG31-235003}, thus violating the reverse isoperimetric inequality
\cite{PRD84-024037}. Therefore, it is very interesting to explore the physics behind these
peculiar properties of the Lorentzian NUT-charged spacetimes.

Very recently, by treating the black hole solutions as topological defects and using the generalized
off-shell free energy, Wei \textit{et al.} \cite{PRL129-191101} constructed the topological numbers
which are independent of the intrinsic parameters of black holes and divided some black hole solutions
into three categories according to their different topological numbers. Because of its simplicity
and easy maneuverability of the procedure, the thermodynamic topological approach proposed in Ref.
\cite{PRL129-191101} soon attracted a great deal of attention and was then successfully applied to
calculate the topological numbers of other black hole solutions \cite{PRD105-104053,PLB835-137591,
PRD106-064059,2208.10177,2211.05524,PRD107-024024,2211.15534,2212.04341,PRD107-084002}. It is natural
to investigate the topological numbers of the NUT-charged spacetimes so as to decree whether they
are black holes or not. In this paper, we shall investigate the topological numbers for the
four-dimensional neutral Lorentzian Taub-NUT, Taub-NUT-AdS and Kerr-NUT spacetimes
by first employing the uniformly modified form of the generalized off-shell Helmholtz free energy,
and find that from the thermodynamic topological standpoint, these spacetimes should be viewed
as generic black holes also.

The organization of this paper is outlined as follows. In Sec. \ref{II}, we give a brief review of
the thermodynamic topological approach proposed in Ref. \cite{PRL129-191101}. In
Sec. \ref{III}, we first recall the consistent formulation of thermodynamic properties of the
four-dimensional static Lorentzian Taub-NUT spacetime and then investigate its topological number.
In Sec. \ref{IV}, we turn to discuss the case of the rotating Lorentzian Kerr-NUT spacetime. In
Sec. \ref{V}, we then extend to discuss the more general static Lorentzian  Taub-NUT-AdS$_4$
spacetime. Finally, our conclusion and outlook are given in Sec. \ref{VI}.

\section{A brief review of thermodynamic topological approach}\label{II}

Following the thermodynamic topological approach proposed in Ref. \cite{PRL129-191101}, one can
first introduce the generalized off-shell Helmholz free energy
\be\label{FE}
\mathcal{F} = M -\frac{S}{\tau}
\ee
for a black hole thermodynamical system with the mass $M$ and the entropy $S$, where $\tau$ is an
extra variable that can be regarded as the inverse temperature of the cavity surrounding the black
hole. Only when $\tau = T^{-1}$, the generalized Helmholz free energy (\ref{FE}) is on-shell and
reduces to the black hole's ordinary Helmholtz free energy $F = M -TS$ \cite{PRD33-2092,PRD105-084030,
PRD106-106015}.

In Ref. \cite{PRL129-191101}, a key vector $\phi$ is defined as
\bea\label{vector}
\phi = \Big(\frac{\p \mathcal{F}}{\p r_{h}}\, , ~ -\cot\Theta\csc\Theta\Big) \, ,
\eea
where the two parameters obey $0 < r_h < +\infty$, $0 \le \Theta \le \pi$, respectively. The
component $\phi^\Theta$ is divergent at $\Theta = 0$ and $\Theta = \pi$, implying that the
direction of the vector is outward here.

A topological current can be defined using the Duan's $\phi$-mapping topological current theory
\cite{SS9-1072,SLAC-PUB-3301,NPB514-705,PRD61-045004} as follows:
\be\label{jmu}
j^{\mu}=\frac{1}{2\pi}\epsilon^{\mu\nu\rho}\epsilon_{ab}\p_{\nu}n^{a}\p_{\rho}n^{b}\, , \qquad
\mu,\nu,\rho=0,1,2,
\ee
where $\p_{\nu}= \p/\p x^{\nu}$ and $x^{\nu}=(\tau,~r_h,~\Theta)$. The unit vector $n$ reads
as $n = (n^r, n^\Theta)$, where $n^r = \phi^{r_h}/||\phi||$ and $n^\Theta
= \phi^{\Theta}/||\phi||$. Since it is easy to demonstrate that the aforementioned current
(\ref{jmu}) is conserved, and one can quickly arrive at $\p_{\mu}j^{\mu} = 0$ and then
demonstrate that the topological current is a $\delta$-function of the field configuration
\cite{NPB514-705,PRD61-045004,PRD102-064039}
\be
j^{\mu}=\delta^{2}(\phi)J^{\mu}\Big(\frac{\phi}{x}\Big) \, ,
\ee
where the three dimensional Jacobian $J^{\mu}(\phi/x)$ satisfies: $\epsilon^{ab}J^{\mu}(\phi/x)
= \epsilon^{\mu\nu\rho}\p_{\nu}\phi^a\p_{\rho}\phi^b$. It is simple to show that $j^\mu$ equals
to zero only when $\phi^a(x_i) = 0$, and one can deduce the topological number $W$ as follows:
\be
W = \int_{\Sigma}j^{0}d^2x = \sum_{i=1}^{N}\beta_{i}\eta_{i} = \sum_{i=1}^{N}w_{i}\, ,
\ee
where $\beta_i$ is the positive Hopf index counting the number of the loops of the vector $\phi^a$
in the $\phi$-space when $x^{\mu}$ are around the zero point $z_i$, while $\eta_{i}= \mathrm{sign}
(J^{0}({\phi}/{x})_{z_i})=\pm 1$ is the Brouwer degree, and $w_{i}$ is the winding number for the
$i$-th zero point of $\phi$ that is contained in the domain $\Sigma$.

\section{Four-dimensional Lorentzian Taub-NUT spacetime}\label{III}

As the simplest case, we will investigate the four-dimensional Lorentzian Taub-NUT spacetime
solution \cite{JMP4-915}, and adopt the following line element in which the Misner strings
\cite{JMP4-924} are symmetrically distributed along the polar axis:
\be\label{NUT}
ds^2 = -\frac{f(r)}{r^2 +n^2}(dt +2n\cos\theta\, d\varphi)^2 +\frac{r^2 +n^2}{f(r)}dr^2
+(r^2 +n^2)(d\theta^2 +\sin^2\theta\, d\varphi^2) \, ,
\ee
where $f(r) = r^2 -2mr -n^2$, in which $m$ and $n$ are the mass and NUT charge parameters,
respectively. The event horizon radius $r_h$ is the largest root of the equation $f(r_h) = 0$,
namely, $r_h = m + \sqrt{m^2 +n^2}$.

\subsection{Consistent thermodynamics}

First, let's briefly recapitulate the ($\psi-\mathcal{N}$)-pair formalism
\cite{PRD100-064055,CQG36-194001} of the consistent thermodynamic of the four-dimensional
Lorentzian Taub-NUT spacetime.

The Bekenstein-Hawking entropy is taken as one quarter of the area of the
event horizon
\be
S = \frac{\mathcal{A}}{4} = \pi(r_h^2 +n^2) \, ,
\ee
and the Hawking temperature is thought of as being proportional to the surface gravity
$\kappa$ on the event horizon
\be
T = \frac{\kappa}{2\pi} = \frac{f^\prime(r_h)}{4\pi(r_h^2 +n^2)}
 = \frac{r_h -m}{2\pi(r_h^2 +n^2)} = \frac{1}{4\pi{}r_h} \, ,
\ee
in which a prime denotes the partial derivative with respective to its variable.

In the ($\psi-\mathcal{N}$)-pair formalism, one intentionally divides the Komar
integral into three patches: the spatial infinity, the horizon, and two Misner string tubes,
and defines a non-globally conserved Misner charge. For the global conserved charge, the Komar
mass related to the timelike Killing vector $\p_t$ is: $M = m$, which is computed via the Komar
integral at infinity.

There are also other Killing horizons associated with the Misner strings in the
Lorentzian Taub-NUT spacetime, namely, the north/south pole axis is a Killing horizon related
to the Killing vector \cite{CQG36-194001}: $k = \p_t +\p_\varphi/(2n)$, whose corresponding
surface gravity can be calculated as
\be
\hat{\kappa} = \frac{1}{2n} \, ,
\ee
and the associated Misner potential is
\be\label{psi}
\psi = \frac{\hat{\kappa}}{4\pi} = \frac{1}{8\pi n} \, .
\ee
It is a simple matter to check that the above thermodynamic quantities simultaneously fulfil
both the differential and integral mass formulae:
\bea
dM &=& TdS +\psi{}d\mathcal{N} \, , \\
M &=& 2TS +2\psi{}\mathcal{N} \, ,
\eea
with the gravitational Misner charge
\be
\mathcal{N} = -\frac{4\pi n^3}{r_h}\, ,
\ee
being conjugate to the Misner potential $\psi$.

The expression of the Helmholtz free energy is then given as \cite{PRD100-064055,CQG36-194001,
PRD105-124034}
\be\label{feTN}
F = M -TS -\psi\mathcal{N} = \frac{m}{2} \, .
\ee

\subsection{Topological number}

Next, we will obtain the topological number of the four-dimensional Lorentzian Taub-NUT spacetime.
The expression for the Helmholtz free energy of the Lorentzian Taub-NUT spacetime can be recovered
via a Wick-rotated back procedure from the Euclidean action of the Euclidean Taub-NUT spacetime:
\be\label{Euact1}
I_E = \frac{1}{16\pi}\int_M d^4x \sqrt{g}R +\frac{1}{8\pi}\int_{\p M} d^3x \sqrt{h}(K -K_0) \, ,
\ee
where $h$ is the determinant of the induced metric $h_{ij}$, $K$ is the trace of the extrinsic
curvature tensor defined on the boundary with this metric, and $K_0$ is the subtracted one of
the massless Taub-NUT solution as the reference background.

The free energy calculated by the action integral is: $I/\beta = m/2 = F$,
where $\beta = 4\pi{}r_h$ is the interval of the time coordinate. Replacing $T$ with $1/\tau$
in Eq. (\ref{feTN}) and using $m = (r_h^2 -n^2)/(2r_h)$, the generalized off-shell Helmholtz
free energy is modified as
\be
\mathcal{F} = M -\frac{S}{\tau} -\psi\mathcal{N} = \frac{r_h}{2} - \frac{\pi(r_h^2 +n^2)}{\tau}\, .
\ee
Utilizing the definition of Eq. (\ref{vector}), the components of the vector $\phi$ can be
easily computed as follows:
\be
\phi^{r_h} = \frac{1}{2} -\frac{2\pi r_h}{\tau} \, , \qquad
\phi^{\Theta} = -\cot\Theta\csc\Theta \, .
\ee
By solving the equation: $\phi^{r_h} = 0$, one can obtain a curve on the $r_h-\tau$ plane,
which is just a straight line for the four-dimensional Taub-NUT spacetime:
\be
\tau = 4\pi{}r_h \, .
\ee

\begin{figure}[h]
\centering
\includegraphics[width=0.45\textwidth]{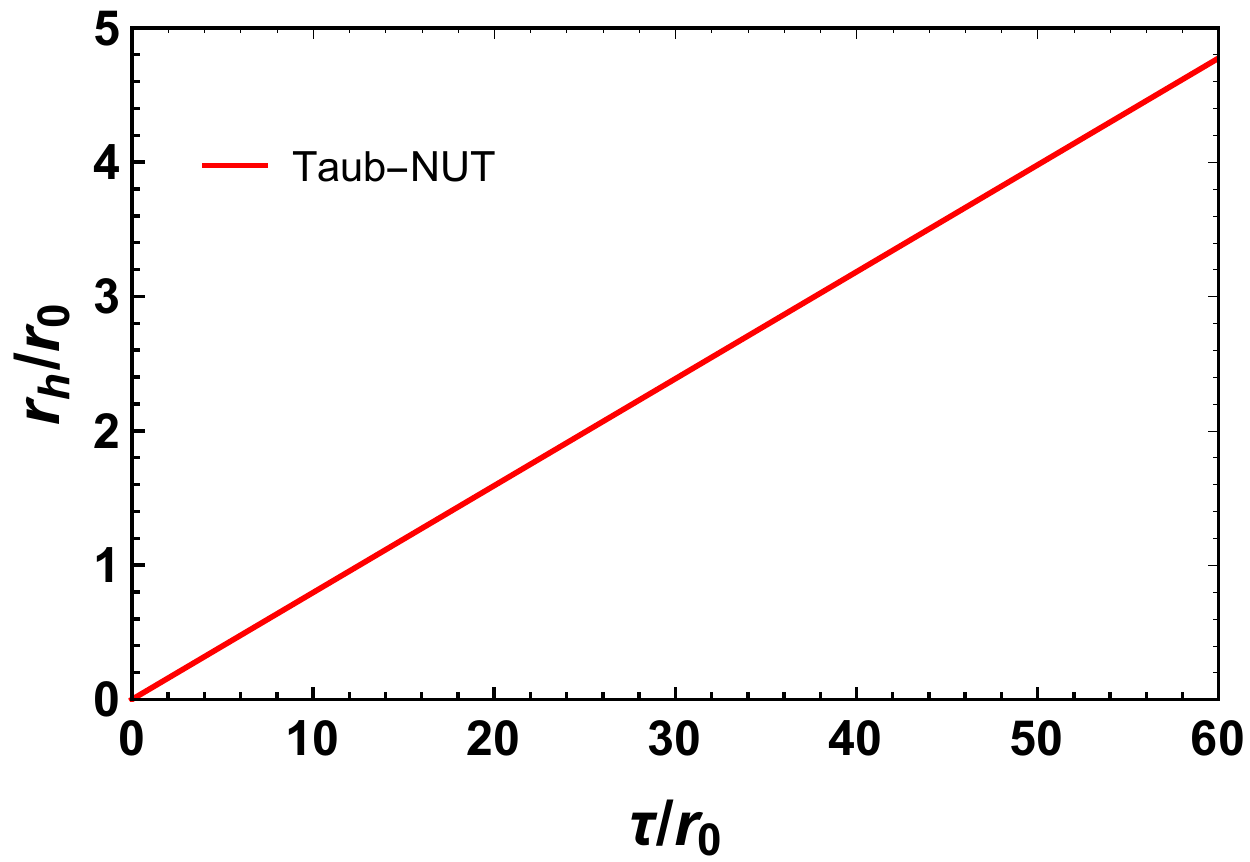}
\caption{Zero points of the vector $\phi^{r_h}$ shown in the $r_h-\tau$ plane.
There is only one Taub-NUT spacetime for any value of $\tau$. \label{4dTaubNUT}}
\end{figure}

\begin{figure}[h]
\centering
\includegraphics[width=0.45\textwidth]{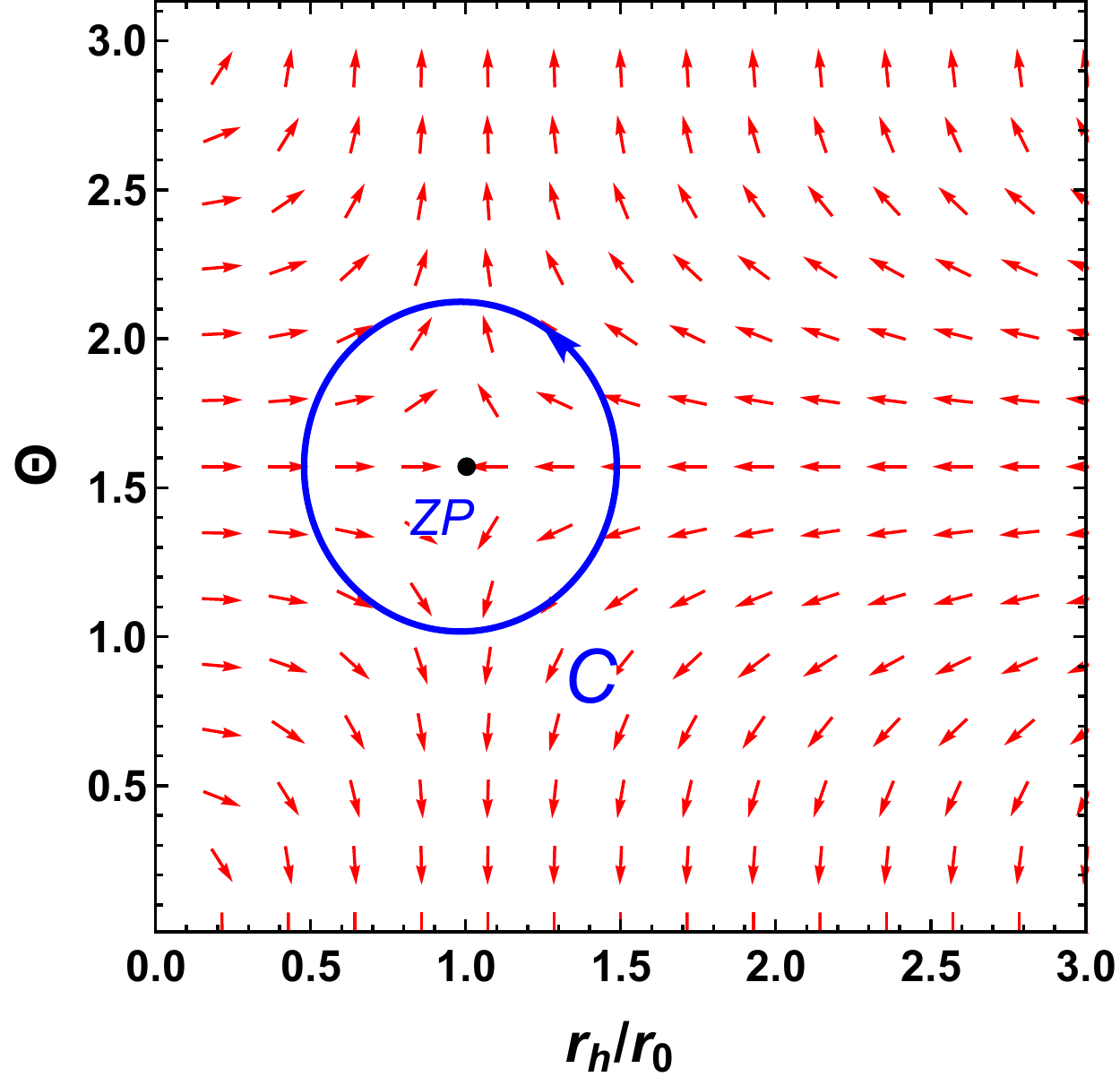}
\caption{The red arrows represent the unit vector field $n$ on a portion of the $r_h-\Theta$
plane for the Taub-NUT spacetime with $\tau/r_0 = 4\pi$. The zero point (ZP) marked with black
dot is at $(r_h/r_0, \Theta) = (1,\pi/2)$. The blue contour $C$ is a closed loop enclosing that
zero point.
\label{TaubNUT4d}}
\end{figure}

For the Lorentzian Taub-NUT spacetime, we plot, respectively, the zero points of the component
$\phi^{r_h}$ in Fig. \ref{4dTaubNUT}, and the unit vector field $n$ on a portion of the $\Theta-r_h$
plane in Fig. \ref{TaubNUT4d} with $\tau = 4\pi r_0$ in which $r_0$ is an arbitrary length scale
set by the size of a cavity enclosing the Taub-NUT spacetime. From Fig. \ref{4dTaubNUT}, one can
observe that the behavior of the Taub-NUT spacetime resembles that of the Schwarzschild black
hole, and this indicates that the NUT charge parameter seems to have no effect on the thermodynamic
topological classification for neutral static asymptotically locally flat spacetime.
Therefore, it would be interesting to explore the connection between the geometric topology and the
thermodynamic topology, for instance, to investigate the topological number of the ultraspinning
black holes \cite{PRD89-084007,PRL115-031101,JHEP0114127,PRD103-104020,PRD101-024057,PRD102-044007,
PRD103-044014,JHEP1121031} and their usual counterparts.

In Fig. \ref{TaubNUT4d}, the zero point is located at $r_h = r_0$ and $\Theta = \pi/2$. one can
determine the winding number $w$ for an arbitrary loop since it is independent of the loops that
surround the zero point; for instance, through a look at the blue contour $C$ in Fig. \ref{TaubNUT4d}.
If the calculation is made for the forward rotation in the anticlockwise direction, then the winding
number is $w = -1$, which coincides with those of the Schwarzschild black hole \cite{PRL129-191101}
and the $d \ge 6$ singly rotating Kerr black hole \cite{PRD107-024024}. In this sense, the Lorentzian
Taub-NUT spacetime behaves like a genuine black hole. Since the winding number is
related to local thermodynamic stability, with positive and negative values corresponding to stable
and unstable black hole solutions \cite{PRL129-191101} respectively, one can naturally conclude
that the Taub-NUT spacetime is an unstable black hole in an arbitrary given temperature just like
the Schwarzschild black hole. Turning to the topological global properties, we have the topological
number $W = -1$ for the Taub-NUT spacetime from Fig. \ref{TaubNUT4d}, which is also the same as
the results of the Schwarzschild black hole and the $d \ge 6$ singly rotating Kerr black hole.
Therefore, from the point of view of topological numbers, the Lorentzian Taub-NUT spacetime should
be accepted into the black hole family. In addition, it can be concluded that although the Taub-NUT
spacetime and Schwarzschild black hole are obviously different in geometric topology aspect, they
are the same class from the perspective of the thermodynamic topology.

\section{Lorentizan Kerr-NUT spacetime}\label{IV}

In this section, we will extend the above discussion to the case of a rotating NUT-charged spacetime
by considering the four-dimensional Kerr-NUT spacetime \cite{DN,CMP10-280,JMP10-1195,JMP14-486},
whose line element with the Misner strings symmetrically distributed along the rotation axis is
\bea\label{KerrNUT}
ds^2 &=& -\frac{\Delta(r)}{\Sigma}\big[dt +(2n\cos\theta -a\sin^2\theta)d\varphi\big]^2
 +\frac{\Sigma}{\Delta(r)}dr^2 \nn \\
&&+\Sigma d\theta^2 +\frac{\sin^2\theta}{\Sigma}\big[adt -(r^2+a^2+n^2)d\varphi\big]^2 \, ,
\eea
where
\be
\Sigma = r^2 +(n +a\cos\theta)^2 \, , \qquad \Delta(r) = r^2 -2mr -n^2 +a^2 \, . \nn
\ee
in which $m$ and $a$ are the mass and rotation parameters, respectively. The horizon is
determined by: $\Delta(r_h) = 0$, which gives $r_h = m \pm\sqrt{m^2 +n^2 -a^2}$.

\subsection{Consistent thermodynamics}

Now we briefly recall the ($\psi-\mathcal{N}$)-pair formalism \cite{PLB798-134972}
of the consistent thermodynamics of the four-dimensional Lorentzian Kerr-NUT spacetime. The
Bekenstein-Hawking entropy is taken as one quarter of the event horizon area:
\be
S = \frac{\mathcal{A}}{4} = \pi(r_h^2 +n^2 +a^2) \, ,
\ee
while the Hawking temperature is proportional to the surface gravity $\kappa$ on
the event horizon
\be
T = \frac{\kappa}{2\pi} = \frac{f^\prime(r_h)}{4\pi(r_h^2 +n^2 +a^2)}
 = \frac{r_h -m}{2\pi(r_h^2 +n^2 +a^2)} \, .
\ee
The angular velocity at the event horizon and the Misner potential are given by
\be
\Omega = \frac{a}{r_h^2 +n^2 +a^2} \, ,
\qquad \psi = \frac{1}{8\pi n} \, .
\ee
As for the global conserved charge, the Komar mass $M = m$ at infinity is related to the
timelike Killing vector $\p_t$.

It is easy to check that the above thermodynamic quantities satisfy the
differential first law and integral Bekenstein-Smarr mass formula simultaneously,
\bea
dM &=& TdS +\Omega{}dJ +\psi{}d\mathcal{N} \, , \\
M &=& 2TS  +2\Omega{}J +2\psi{}\mathcal{N}  \, ,
\eea
with the gravitational Misner charge and the angular momentum
\be
\mathcal{N} = -\frac{4\pi{}n^3}{r_h} \, , \qquad
J = \Big(m +\frac{n^2}{r_h} \Big)a \, ,
\ee
being conjugate to the Misner potential $\psi$ and the angular velocity, respectively. Note
that both of them do not have a global character, due to having a location-dependent
factor $1/r_h$.

With the help of the above expressions and using $m = (r_h^2 -n^2 +a^2)/(2r_h)$,
the Gibbs free energy of the Kerr-NUT spacetime reads
\be\label{GEKN}
G \equiv M -TS -\psi\mathcal{N} -\Omega{}J = \frac{m}{2} \, ,
\ee
which is identical to the one calculated via the action integral \cite{PLB798-134972},
just like the non-rotating case.

\bigskip
\subsection{Topological number}

In order to calculate the thermodynamical topological number of the Kerr-NUT
spacetime, we then consider the Helmholtz free energy which is given by
\be
F = G +\Omega{}J = M -TS -\psi\mathcal{N} \, .
\ee
It coincides with the result of Eq. (4.19) given in Ref. \cite{JHEP0520084} in the case of the
parameters $e = g = 0$ are turned off.

It is straightforward to modify the generalized Helmholtz free energy as
\be
\mathcal{F} = M -\frac{S}{\tau} -\psi\mathcal{N}
 = \frac{r_h^2 +a^2}{2r_h} -\frac{\pi(r_h^2 +n^2 +a^2)}{\tau} \, .
\ee
Then, the components of the vector $\phi$ are given by
\be
\phi^{r_h} = \frac{r_h^2 -a^2}{2r_h^2} -\frac{2\pi r_h}{\tau} \, , \qquad
\phi^{\Theta} = -\cot\Theta\csc\Theta \, .
\ee
Therefore, using the thermodynamic topological approach and solving the equation: $\phi^{r_h} = 0$,
we get
\be
\tau = \frac{4\pi r_h^3}{r_h^2 -a^2}
\ee
as the zero point of the vector field $\phi^{r_h}$.

\begin{figure}[h]
\centering
\includegraphics[width=0.45\textwidth]{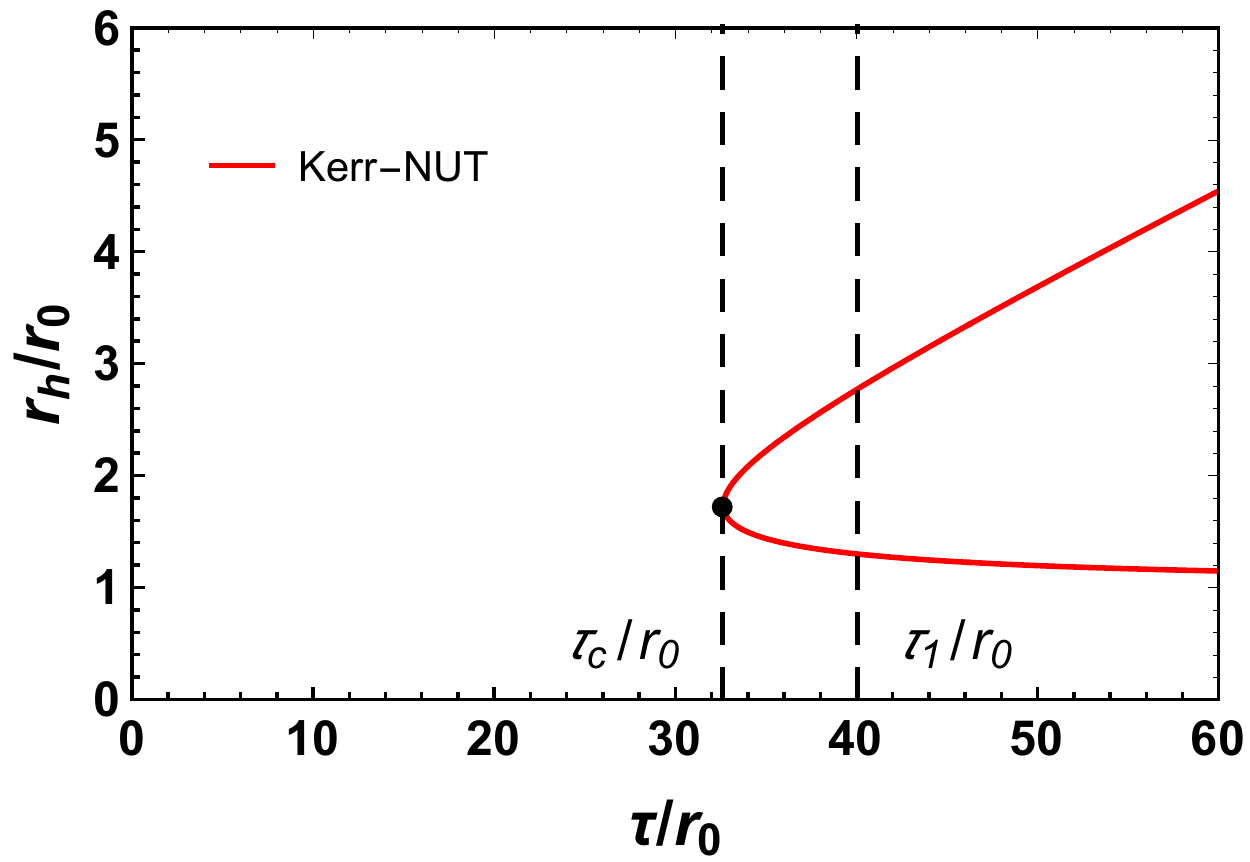}
\caption{Zero points of the vector $\phi^{r_h}$ shown in the $r_h-\tau$ plane.
The generation point for the Kerr-NUT spacetime is represented by the black dot
with $\tau_{c}$. At $\tau = \tau_1$, there are two Kerr-NUT spacetimes. Obviously,
the topological number is: $W=-1+1=0$.} \label{4dKerrNUT}
\end{figure}

\begin{figure}[t]
\centering
\includegraphics[width=0.45\textwidth]{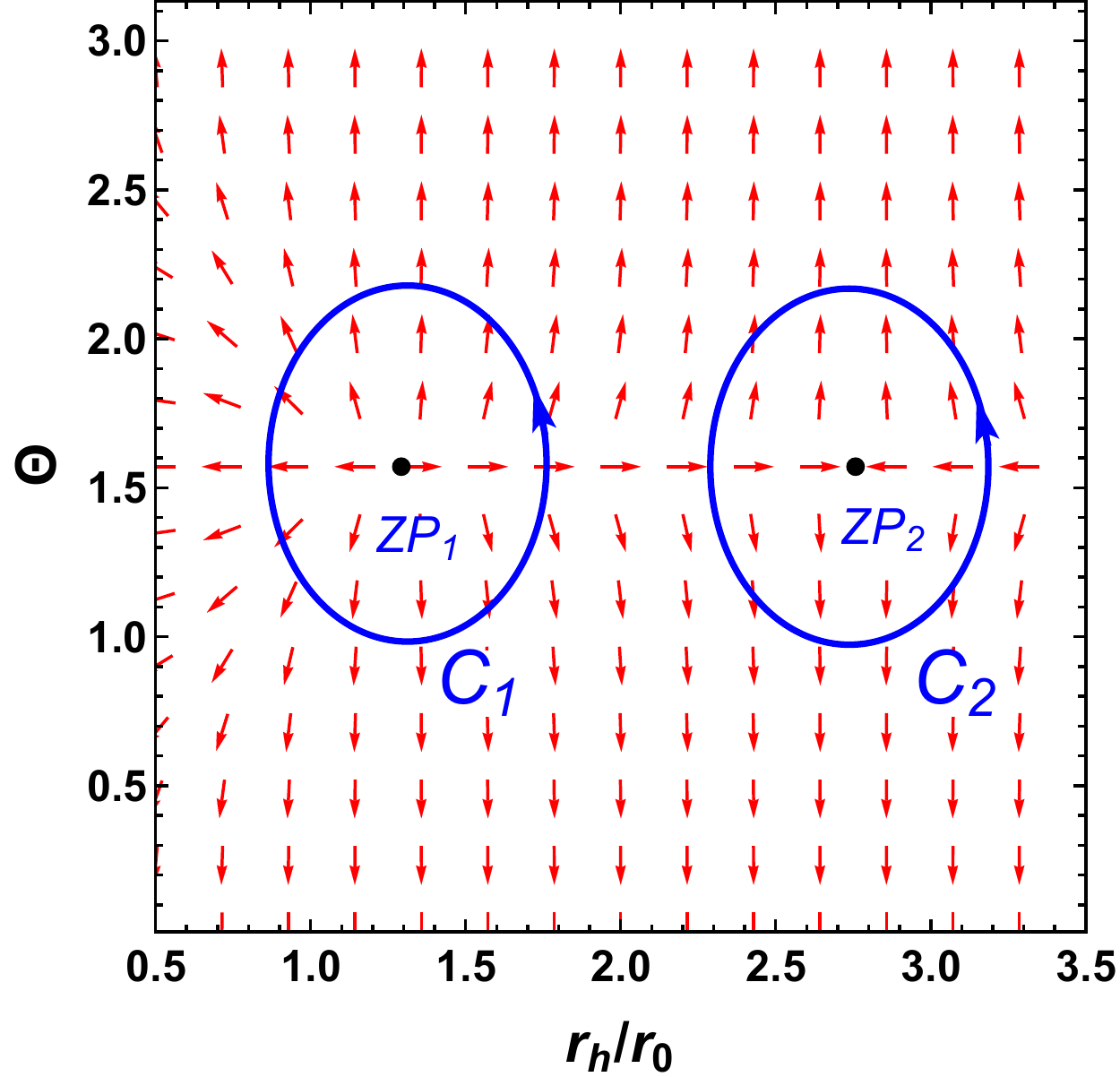}
\caption{The red arrows represent the unit vector field $n$ on a portion of the
$r_h-\Theta$ plane for the Kerr-NUT spacetime with $a/r_0=1$ and $\tau/r_0 = 40$.
The zero points (ZPs) marked with black dots are at $(r_h/r_0, \Theta) = (1.30,\pi/2)$,
and $(2.77,\pi/2)$ for ZP$_1$ and ZP$_2$, respectively. The blue contours $C_i$ are
closed loops surrounding the zero points.} \label{KerrNUT4d}
\end{figure}

Taking $a = r_0$ for the Kerr-NUT spacetime, we plot the zero points of the component $\phi^{r_h}$
in Fig. \ref{4dKerrNUT}, and the unit vector field $n$ on a portion of the $\Theta-r_h$ plane in Fig.
\ref{KerrNUT4d} with $\tau/r_0 = 40$, respectively. In Fig. \ref{4dKerrNUT}, one generation point can
be found at $\tau/r_0 = \tau_{c}/r_0 = 32.65$. In Fig. \ref{KerrNUT4d}, the zero points are located
at $(r_h/r_0, \Theta) = (1.30,\pi/2)$, and $(2.77,\pi/2)$, respectively. Based upon the local property
of the zero points, we can obtain the topological number: $W = -1 +1 = 0$ for the Kerr-NUT spacetime,
which is same as that of the Kerr black hole \cite{PRD107-024024}. Thus, the big family of the black
holes should contain the four-dimensional Kerr-NUT spacetime. In addition, it also can be concluded
that although the Kerr-NUT spacetime and Kerr black hole are obviously different in geometric topology
aspect, they seem to be the same kind from the perspective of the thermodynamic topology, just like
the Taub-NUT spacetime and Schwarzschild black hole as discussed in Sec. \ref{III} and Ref.
\cite{PRL129-191101}, respectively.

\section{Lorentzian Taub-NUT-AdS$_4$ spacetime}\label{V}

In this section, we turn to explore the Lorentzian Taub-NUT spacetime with an negative
cosmological constant, namely, the Lorentzian Taub-NUT-AdS$_4$ spacetime, whose metric is
still given by Eq. (\ref{NUT}), but now $f(r) = r^2 -2mr -n^2 +(r^4 +6n^2r^2 -3n^4)/l^2$,
in which the AdS radius $l$ appears in the thermodynamic pressure $P = 3/(8\pi{}l^2)$ of
the 4-dimensional AdS black hole \cite{PRD84-024037,CQG26-195011}.

\subsection{Consistent thermodynamics}

We now recollect the thermodynamical properties of the four-dimensional Taub-NUT-AdS
spacetime. The Bekenstein-Hawking entropy is taken as one quarter of the event horizon area:
\be
S = \frac{\mathcal{A}}{4} = \pi(r_h^2 +n^2) \, ,
\ee
and the Hawking temperature is assumed to be proportional to the surface gravity $\kappa$ on the
event horizon
\be
T = \frac{\kappa}{2\pi} = \frac{f^\prime(r_h)}{4\pi(r_h^2 +n^2)} = \frac{r_h -m}{2\pi(r_h^2 +n^2)}
+\frac{r_h^3 +3n^2r_h}{\pi{}l^2(r_h^2 +n^2)} \, ,
\ee
where $r_h$ is the location of the event horizon.

The conformal mass is: $M = m$, and the Misner potential is
\be
\psi = \frac{1}{8\pi n} \, .
\ee
It is easy to verify that the above thermodynamic quantities satisfy the differential first law and
integral Bekenstein-Smarr mass formula simultaneously \cite{PRD100-064055,CQG36-194001},
\bea
dM &=& TdS +\psi{}d\mathcal{N} +VdP \, , \\
M &=& 2TS +2\psi{}\mathcal{N} -2VP \, ,
\eea
with the gravitational Misner charge and the thermodynamic volume
\be
\mathcal{N} = -\frac{4\pi n^3}{r_h}\Big[1 +\frac{3(n^2 -r_h^2)}{l^2} \Big] \, , \qquad
V = \frac{4}{3}\pi r_h(r_h^2 +3n^2) \, ,
\ee
being conjugate to the Misner potential $\psi$ and the pressure $P = 3/(8\pi l^2)$, respectively.

\subsection{Topological number}

One can find that the Helmholtz free energy reads \cite{PRD100-064055}
\be
F \equiv M -TS -\psi\mathcal{N} = \frac{m}{2} -\frac{r_h(r_h^2 +3n^2)}{2l^2} \, ,
\ee
which coincides with those computed via the action integral, namely $F = I/\beta$. In order to
get this result, one can calculate the Euclidean action \cite{PRD100-064055} for the Euclidean
spacetime
\be
I_E = \frac{1}{16\pi}\int_M d^4x \sqrt{g}\Big(R +\frac{6}{l^2}\Big) +\frac{1}{8\pi}\int_{\p M} d^3x\sqrt{h}K
 -\frac{1}{8\pi}\int_{\p M} d^3x \sqrt{h}\Big[\frac{2}{l} +\frac{l}{2}\mathcal{R}(h)\Big] \, ,
\ee
where $K$ and $\mathcal{R}(h)$ are the extrinsic curvature and Ricci scalar of the boundary metric,
respectively. In order to cancel the divergence, the above action also contains the Gibbons-Hawking
boundary term and the necessary AdS boundary counterterms \cite{PRD60-104001,PRD60-104026,PRD60-104047,
CMP208-413,CMP217-595}, apart from the standard Einstein-Hilbert term.

It is now a position to investigate the topological number of the four-dimensional
Lorentz Taub-NUT-AdS spacetime. Replacing $T$ with $1/\tau$ and substituting $l^2 = 3/(8\pi P)$,
thus the modified form of the generalized off-shell Helmholtz free energy is
\be
\mathcal{F} = M -\frac{S}{\tau} -\psi\mathcal{N} =
\frac{r_h}{2} +\frac{4\pi P}{3}r_h(r_h^2 +3n^2) -\frac{\pi(r_h^2 +n^2)}{\tau} \, .
\ee
Then, the components of the vector $\phi$ are obtained as follows:
\be
\phi^{r_h} = \frac{1}{2} +4\pi P(r_h^2 +n^2) -\frac{2\pi r_h}{\tau} \, ,  \qquad
\phi^{\Theta} = -\cot\Theta\csc\Theta \, .
\ee
from which one can get the zero point of the vector field $\phi^{r_h}$ as
\be\label{tauTNA}
\tau = \frac{4\pi r_h}{8\pi P(r_h^2 +n^2) +1} \, .
\ee
Note that Eq. (\ref{tauTNA}) consistently reduces to the one obtained in the four-dimensional
Schwarzschild-AdS black hole case \cite{PRD107-084002} when the NUT charge parameter $n$ vanishes.

\begin{figure}[h]
\centering
\includegraphics[width=0.45\textwidth]{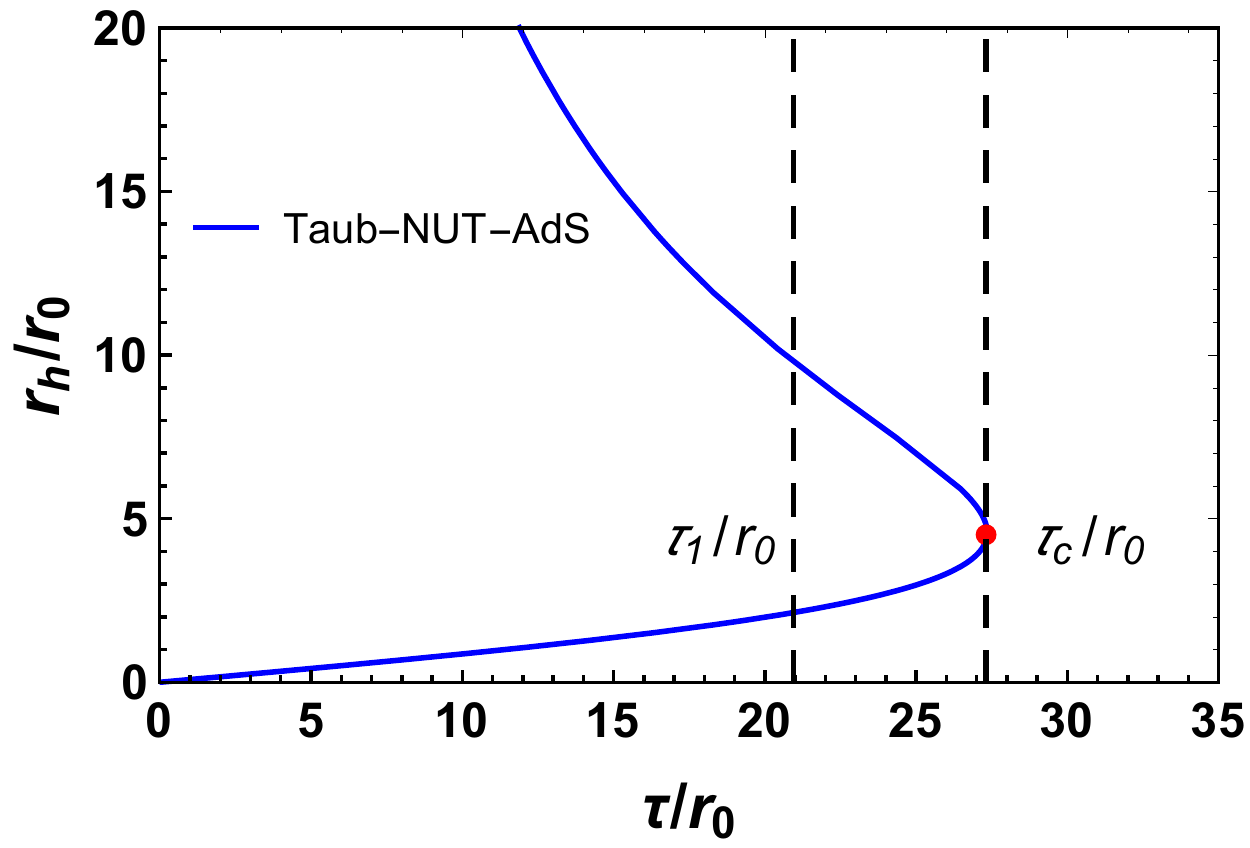}
\caption{
Zero points of the vector $\phi^{r_h}$ shown on the $r_h-\tau$ plane
with $Pr_0^2=0.002$ for the Taub-NUT-AdS$_4$ spacetime. The annihilation
point for this spacetime is represented by the red dot with $\tau_c$. There
are two Taub-NUT-AdS$_4$ spacetimes when $\tau = \tau_1$. Obviously, the
topological number is: $W = 1-1 = 0$.} \label{4dTNAdS}
\end{figure}

\begin{figure}[t]
\centering
\includegraphics[width=0.45\textwidth]{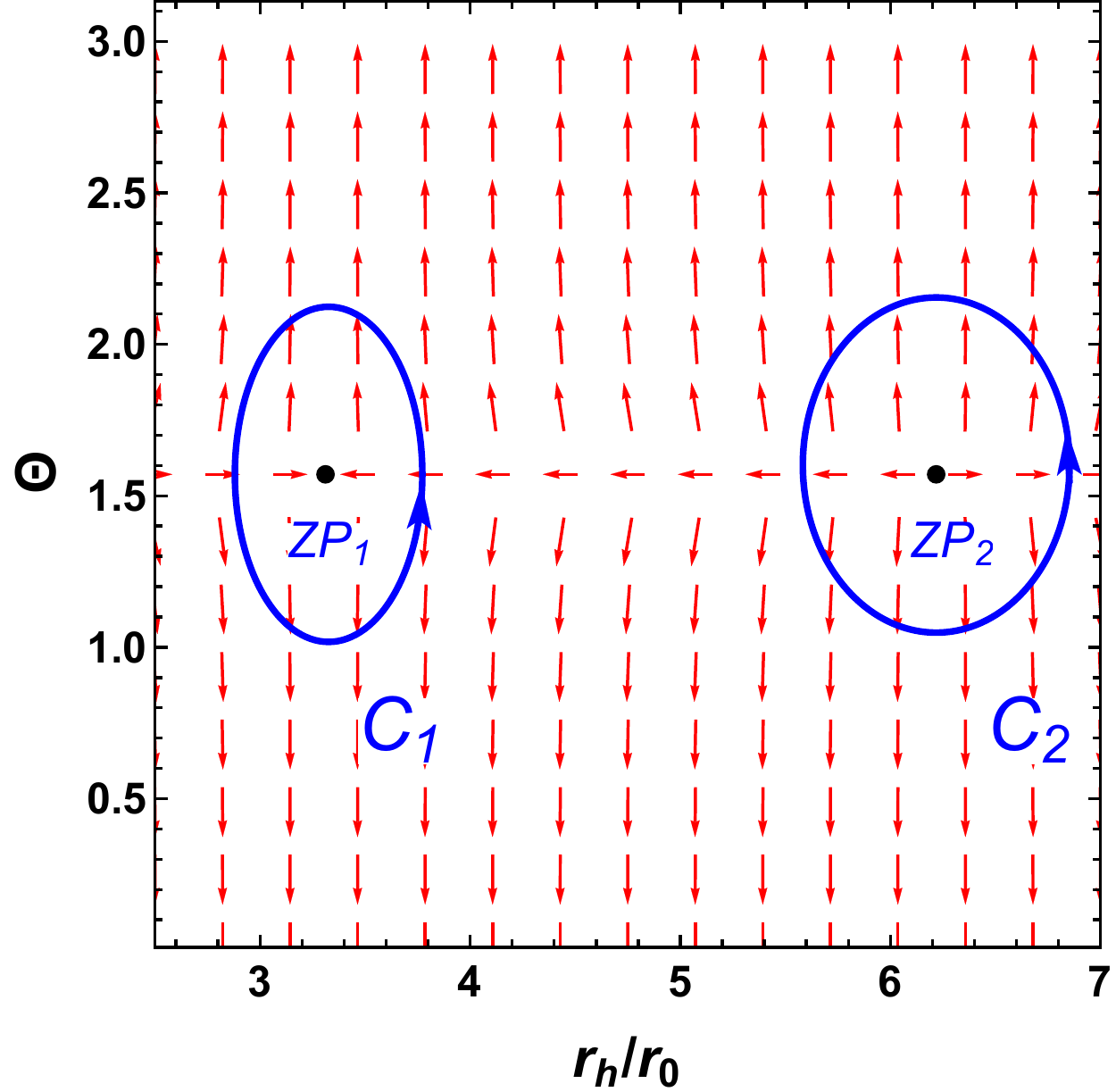}
\caption{The red arrows represent the unit vector field $n$ on a portion
of the $r_h-\Theta$ plane with $Pr_0^2=0.002$ and $\tau/r_0=26$ for the Taub-NUT-AdS$_4$
spacetime. The zero points (ZPs) marked with black dots are at $(r_h/r_0, \Theta) = (3.32,
\pi/2)$, $(6.30, \pi/2)$ for ZP$_1$ and ZP$_2$, respectively. The blue contours $C_i$ are
closed loops surrounding the zero points.}
\label{TNAdS4d}
\end{figure}

Taking the pressure $Pr_0^2 = 0.02$ and the NUT charge parameter $n/r_0 = 1$, in Figs. \ref{4dTNAdS}
and \ref{TNAdS4d}, we plot the zero points of $\phi^{r_h}$ in the $r_h-\tau$ plane and the unit
vector field $n$ on a portion of the $\Theta-r_h$ plane for the Taub-NUT-AdS$_4$ spacetime. For the
Taub-NUT-AdS$_4$ spacetime, we observe that the topological number is $W = 0$, which is the same as
that of the Schwarzschild-AdS$_4$ black hole \cite{PRD106-064059,PRD107-084002}, and this implies that
the NUT charge parameter also seems to have no impact on the thermodynamic topological classification
for the neutral static asymptotically local AdS spacetime. Furthermore, it indicates that the
Taub-NUT-AdS$_4$ solution can still be categorized as one of the three types of known black hole
solutions \cite{PRL129-191101}. As a result, at least according to the thermodynamic topological
approach, the Lorentzian Taub-NUT-AdS$_4$ spacetime should be included into a member of
the black hole family.

\section{Conclusion and outlook}\label{VI}

\begin{table}[h]
\caption{The topological number $W$, numbers of generation and annihilation points
for the four-dimensional neutral Lorentzian NUT-charged spacetimes.}
\begin{tabular}{c|c|c|c}
\hline\hline
Solutions & $W$ & Generation point & Annihilation point\\ \hline
Taub-NUT & -1 & 0 & 0\\
Kerr-NUT & 0 & 1 & 0\\
Taub-NUT-AdS & 0 & 0 & 1\\
\hline\hline
\end{tabular}
\label{I}
\end{table}

Our results obtained in this paper are summarized in Table \ref{I}. In this work, we have
employed the uniformly modified form of the generalized off-shell Helmholtz free energy and investigated
the topological numbers for the four-dimensional uncharged Lorentzian Taub-NUT, Taub-NUT-AdS and Kerr-NUT
spacetimes. We showed that the Taub-NUT spacetime has: $W = -1$, which is the same as that of the Schwarzschild
black hole \cite{PRL129-191101}. We found that the Kerr-NUT spacetime has: $W = 0$, which is identical to that
of the Kerr black hole \cite{PRD107-024024}. We also demonstrated that the Taub-NUT-AdS$_4$ spacetime has:
$W = 0$, which coincides with that of the Schwarzschild-AdS$_4$ black hole \cite{PRD107-084002}. Therefore,
one can conclude that the existence of the NUT charge parameter seems has no impact on the topological
number of the neutral asymptotically locally flat/AdS spacetimes. It can be inferred that the
four-dimensional Taub-NUT, Taub-NUT-AdS and Kerr-NUT spacetimes should be viewed as generic
black holes from the viewpoint of the thermodynamic topological approach.

There are two promising topics to be pursued in the future. The first intriguing
topic is to explore the phase transitions of the NUT-charged AdS spacetimes, such as the Hawking-Page
phase transitions \cite{CMP87-577} and the P-V criticality \cite{JHEP0712033}, and we expect that
this will help to unravel the mystery of the NUT-charged spacetimes. The second interesting topic
is to extend the present work to the more general charged RN-NUT(-AdS) and rotating charged
Kerr-Newman-NUT cases.

Perhaps, one has just touched the tip of an iceberg, much more needs to be explored.

\begin{acknowledgments}
We thank Professor Shuang-Qing Wu for helpful suggestions and detailed discussions.
We are also greatly indebted to the anonymous referee for his/her
constructive comments to improve the presentation of this work. This work
is supported by the National Natural Science Foundation of China (NSFC) under Grant No. 12205243,
No. 11675130, by the Sichuan Science and Technology Program under Grant No. 2023NSFSC1347,
by the Sichuan Youth Science and Technology Innovation Research Team (21CXTD0038), and by the
Doctoral Research Initiation Project of China West Normal University under Grant No. 21E028.
\end{acknowledgments}

\end{document}